\documentclass[a4paper,11pt,superscriptaddress]{article}
\pdfoutput=1 
\usepackage{jheppub} 
\usepackage{hyperref}
\usepackage{booktabs}
\usepackage{cleveref}
\usepackage[T1]{fontenc} 
\usepackage{siunitx} 
\usepackage{xcolor,graphicx}
\usepackage{dcolumn}
\usepackage{bm}
\usepackage[normalem]{ulem}
\usepackage{braket} 
\usepackage{amsmath}
\usepackage{cancel}
\usepackage{slashed}
\usepackage{multicol}

\newcommand\be{\begin{equation}}
\newcommand\ee{\end{equation}}

\newcommand\bea{\begin{eqnarray}}
\newcommand\eea{\end{eqnarray}}

\begin{document}
\bibliographystyle{apsrev4-1}

\title{Using DUNE to Shed Light on the Electromagnetic Properties of Neutrinos}

\author[a]{Varun Mathur,}
\author[a]{Ian M. Shoemaker,}
\author[a,b]{Zahra Tabrizi}
\affiliation[a]{Center for Neutrino Physics, Department of Physics, Virginia Tech University, Blacksburg, VA 24601, USA}
\affiliation[b]{Northwestern University, Department of Physics \& Astronomy, 2145 Sheridan Road, Evanston, IL 60208, USA}

\emailAdd{varun@vt.edu,
          shoemaker@vt.edu,
          ztabrizi@northwestern.edu}
\abstract{
We study future DUNE sensitivity to various electromagnetic couplings of neutrinos, including magnetic moments, milli-charges, and charge radii. The DUNE PRISM capabilities play a crucial role in constraining the electron flavored couplings. We find that DUNE will be able to place the strongest terrestrial constraint on the muon-neutrino magnetic moment by improving on LSND's bounds by roughly a factor of two, although Borexino's solar constraint will still be stronger. For the muon neutrino milli-charge DUNE can place the leading experimental bound, with two orders of magnitude improvement compared to the existing COHERENT constraint. Finally, DUNE may be able to test the SM prediction for the muon neutrino charge radius, by placing a constraint two times better than CHARM-II and CCFR experiments. 
}

\maketitle

\section{Introduction}

Exploiting the information that can be gained from neutrino-electron ($\nu-e$) scattering to understand the electromagnetic properties of the neutrinos goes back nearly 90 years, to a 1932 paper by Carlson and Oppenheimer~\cite{Carlson:1932rk} which considered probes of the magnetic properties of the neutrino (then called a ``neutron''). Moreover the notion that neutrinos may have non-trivial electromagnetic properties is as old as the neutrino itself, given that Pauli's original letter in 1930 supposed that neutrinos may have detectable magnetic moments. The purpose of the present paper is to explore the extent to which the upcoming DUNE near detector complex can further this story, and shed light on the electromagnetic properties of neutrinos. 

Of course the most intriguing known property of neutrinos is that they are massive. As such, neutrinos require additional physics beyond the Standard Model (SM) of particle physics. In many such extensions of the SM, neutrinos acquire electromagnetic interactions as a result of loops of charged particles at high mass scales. Furthermore, neutrinos offer a relatively clean experimental probe of new physics since unlike the rest of the SM fermions they do not participate in known electromagnetic or strong nuclear interactions. In most experiments the main neutrino electromagnetic property which has been studied is its magnetic moment, with many stringent bounds from several short and long baseline experiments put on them. However, there are several models beyond the SM that assume neutrinos have a very small electric charge, often called the neutrino ``milli-charge'' (see e.g. \cite{Foot:1989fh,Babu:1992sw}). On the other hand, even if the electric charge of neutrinos are zero, they can still have a finite charge radius. While Refs.~\cite{Broggini:2012df,Giunti:2014ixa} contain a through review of the theoretical and experimental impact of neutrino electromagnetic properties up to around 2014, many additional experimental probes are available today. The general electromagnetic properties of neutrinos in their mass basis is discussed in Ref.~\cite{Kouzakov:2017hbc}. Moreover, the observation of coherent elastic neutrino-nucleus scattering by the COHERENT collaboration has provided new strong bounds on the neutrino milli-charge~\cite{Cadeddu:2020lky} and the transition magnetic moment \cite{Miranda:2019wdy}.  

Although the data of decades of neutrino experiments are in good agreement with the standard picture of neutrinos, there are still unanswered questions, like the CP violating phase or the octant of the $\theta_{23}$ mixing angle. Future experiments like DUNE are planning to answer some of these questions. The near detectors of these experiments are designed to mainly control the systematic uncertainties, but thanks to their large fiducial mass and the extremely large neutrino fluxes that reach to the near detectors, we can use them to search for physics beyond the standard model, or study rare neutrino properties. In this work we use the $\nu-e$ interaction at the DUNE near detector complex to study the electromagnetic properties of neutrinos and get constraints on the  magnetic moment, the neutrino electric milli-charge and the charge radius of neutrinos. We study the sensitivity of DUNE to neutrino electromagnetic properties by using the PRISM concept, a near detector which can move perpendicular with respect to the beam axis.

The remainder of this paper is organized as follows. In Sec.~\ref{sec:sm} we discuss the neutrino-electron scattering predictions from the SM. In Sec.~\ref{sec:DUNE} we discuss the DUNE flux and near detector performance assumptions we make. In Sec.~\ref{sec:EMproperties} we introduce the electromagnetic properties we will focus on, the impact these have on the $\nu-e$ scattering, and our resulting projected sensitivities. Finally in Sec.~\ref{sec:concl} we conclude and discuss possible future extensions of our work. 

\section{Neutrino-electron scattering}
 \label{sec:sm}
 
At the SM the tree-level differential cross-sections for neutrino-electron ($\nu-e$) scattering is given by

\begin{eqnarray}\label{eq:nueXSSM}
\frac{d\sigma^{\rm{SM}}}{dE_R}(\nu_\alpha  e^-\rightarrow \nu_\alpha e^-)&=&\frac{m_e G_f^2 }{2\pi}\Bigg[\Big(g_{V}^{\nu_\alpha }\pm g_{A}^{\nu_\alpha }\Big)^2+\Big(g_{V}^{\nu_\alpha }\mp g_{A}^{\nu_\alpha }\Big)^2\Big(1-\frac{E_R}{E_\nu}\Big)^2\\&-&\Big((g_{V}^{\nu_\alpha })^2-(g_{A}^{\nu_\alpha })^2\Big) \frac{m_eE_R}{E^2_\nu}\Bigg]\,.\nonumber
\end{eqnarray}
 Here $m_e$ and $E_R$ are the mass and the recoil kinetic energy of electron, $E_\nu$ is the incoming neutrino energy and $G_f$ is the Fermi constant. The vector ($g_{V}^{\nu_\alpha }$) and axial ($g_{A}^{\nu_\alpha }$) couplings for neutrinos are:
\begin{eqnarray}\label{eq:VecAxcouplings}
g_{a}^{\nu_\alpha }&=&(g_{a}^{\rm{SM}})^{\nu_\alpha }+\delta_{\alpha e}\,,~~~~~~~~~~~~(a=V,A)\,,
\end{eqnarray}
where $(g_{V}^{\rm{SM}})^{\nu_\alpha }=-\frac{1}{2}+2s^2_w$ and $(g_{A}^{\rm{SM}})^{\nu_\alpha }=-\frac{1}{2}$. The term $\delta_{\alpha e}$ appears because the electron neutrinos interact with both $W$ and $Z$ bosons, while the muon and tau neutrinos only interact with $Z$. We fix the weak mixing angle $s^2_w\equiv \sin^2\theta_w=0.2385$~\cite{Tiesinga:2021myr}, which corresponds to the  average energy scale $\langle Q^2\rangle \sim (50~{\rm MeV})^2$ in the $\overline{\scalebox{0.8}{\text{MS}}}$ scheme relevant for DUNE. Finally, the $-~(+)$ sign in the first (second) term denotes to anti-neutrinos.

Kinematics dictate the following relation between the angle of the scattered electron, $\theta$, and the energy of the outgoing electron, $E_e$:
\bea
1-\cos\theta=m_e\frac{1-y}{E_e},
\eea
where $E_e=m_e+E_R$, $y\equiv E_R/E_\nu$ is the inelasticity with $y_{min}=E_R^{th}/E_\nu$ and $y_{max}=1$, and $E_R^{th}$ is the threshold energy set by the energy resolution of the detector. The signature of a $\nu-e$ scattering event is a forward going electron with small $\theta$, hence $E_e\theta^2<2m_e$ is imposed. This is of course only true for a perfect energy resolution of the detector. In reality the reconstructed and the true electron energies are different and we can have events even for $E_e\theta^2>2m_e$.  Hence, we do a binning with respect to $E_e\theta^2$ with the reconstructed electron energy, which also helps in improving the background rejection (more details in Sec.~\ref{sec:DUNE}). In this case the differential cross section becomes \cite{Ballett:2019xoj}
\begin{eqnarray}\label{eq:nuexsection}
\frac{d\sigma}{d(E_e\theta^2)}=\frac{E_\nu}{2m_e}\frac{d\sigma_{\nu_{\alpha}e}}{dE_R} \Bigg|_{E_R=E_\nu(1-\frac{E_e\theta^2}{2m_e})}\,.
\end{eqnarray}
Finally, using this differential cross section we calculate the expected number of $\nu-e$ events at each bin of $E_e\theta^2$ based on Eq.~(\ref{eq:expEvents}). 
\section{DUNE setup}
\label{sec:DUNE}

\begin{figure}[t!]
\includegraphics[angle=0,width=1\textwidth]{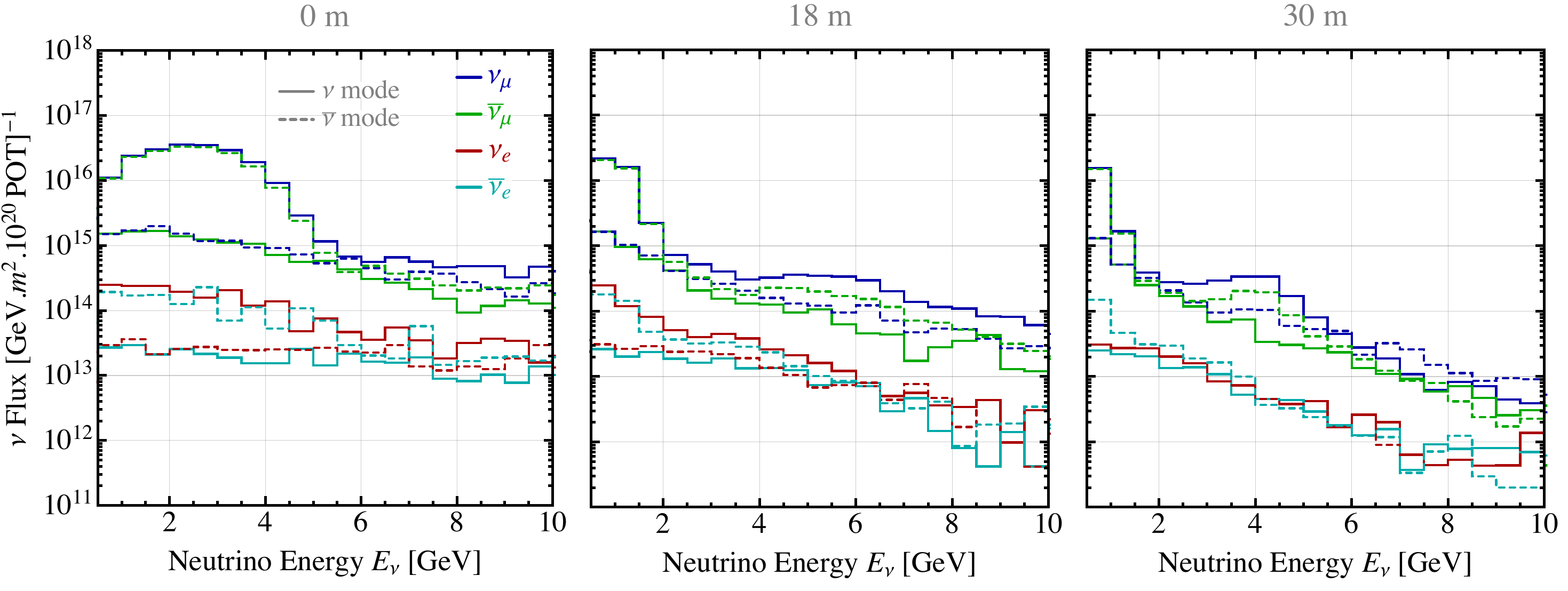}
\caption{The neutrino fluxes in the on-axis and two off-axis positions we have considered in this work, based on ~\cite{DUNE:flux_updated}. For each neutrino (solid) or anti-neutrino (dashed) modes we have shown the fluxes for $\nu_\mu$ (blue), $\bar\nu_\mu$ (green), $\nu_e$ (red) and $\bar\nu_e$ (cyan).
}
\label{fig:fluxes}
\end{figure}


To estimate the number of $\nu-e$ events at the DUNE near detector we assume 50-ton fiducial mass of liquid argon time projection chamber (LArTPC) located 574 m from the source. For the neutrino fluxes we use the intense neutrino beam at the Long-Baseline Neutrino Facility (LBNF) fluxes assuming 1.2-MW, 120-GeV proton beam with $1.1\times10^{21}$ protons on target (POT) per year taken from Ref.~\cite{DUNE:flux_updated}, as described in the DUNE Conceptual Design Report \cite{DUNE:2021tad}. For the lifetime of the experiment we assume two cases: i) The experiment runs for 7 years equally in the neutrino and anti-neutrinos modes ($3.5$ years each), but all the time in the on-axis location (0 m); ii) The experiment runs for $3.5$ years in the on-axis location (0 m), and $3.5$ years in the off-axis locations of 18 m and 30 m of the PRISM concept, equally divided between the neutrino and anti-neutrino modes. We have shown the relevant fluxes for each case in Fig.~\ref{fig:fluxes}.

For the 7-year exposure, The DUNE collaboration expects to observe $4.5\times10^4$ $\nu-e$ events if the experiment is all the time in the on-axis location, or $2.6\times10^4$ events in the PRISM concept. We have shown in Table \ref{tab:LArrates} the expected number of $\nu-e$ events per year for each neutrino flavor at each detector location, for both the neutrino and anti-neutrino modes. The cyan shaded regions in Fig.s \ref{fig:magneticEvents}, \ref{fig:millichargeEvents} and \ref{fig:radiusEvents} show the expected SM number of $\nu-e$ events per year at each bin of $E_e\theta^2$. The LArTPCs can have a great angular resolution $\mathcal{O}(1^\circ)$ for electromagnetic showers~\cite{DUNE:2021tad}. To estimate the expected number of events at each bin we have assumed an angular resolutions of $\sigma_\theta=1^\circ$  and an energy resolution of $\sigma/E = 10\%/\sqrt{E}$. Finally, we have assumed a threshold energy of $E_R^{th}=30$~MeV (see appendix \ref{app:resolution} for details).

Although the neutrino fluxes at DUNE will probably have very large uncertainties, the DUNE-PRISM concept can reduce the flux uncertainties by looking into the ratios of on-axis to off-axis fluxes, for which the uncertainties would be only dictated by better understood meson-decay kinematics. 
On the other hand, although due to the off-axis location of the PRISM detectors the expected number of events decreases for all the neutrino flavors, the ratio of the electron to muon neutrino events increase by going to the off-axis locations (See e.g. Fig.~1 of \cite{deGouvea:2019wav}). Therefore, in PRISM the $\nu_e$  events become more relevant, and we expect to get better constraints on its electromagnetic properties.  

\begin{table}[t]
\begin{center}
\scalebox{0.9}{
\begin{tabular}{|c|c|c|c|c|c|c|c|}
\hline\hline
		\bf Channel & \bf  0 m&  \bf 18 m  &\bf  30 m   \\ \hline \hline
    $\nu_\mu e\to \nu_\mu e$& 5,915 &573 & 175 \\
    & 864 &  161 &  65 \\\hline
    $\bar\nu_\mu e\to \bar\nu_\mu e$& 593 & 102  &  39 \\
    & 4,515 & 409  &  126  \\\hline
   $\nu_e e\to \nu_e e$ & 574 & 108 &39 \\
    &  285 & 54& 23 \\
    \hline
    $\bar\nu_e e\to \bar\nu_e e$ & 78    & 21  &   9  \\
    &   173  &  35 &  13 \\
    \hline\hline
    {\rm{Total}} & 7,160  &  804  & 262 \\
        & 5,837  &  658  &  227 \\
        \hline\hline
\end{tabular}
}
\end{center}
\caption{\label{tab:LArrates} 
The total expected number of $\nu-e$ events per year for different on/off axis locations for each flavor. In each entry the top (bottom) number belongs to the (anti)neutrino mode.
 }
\end{table}

The main background sources for the $\nu-e$ events are the misidentified $\pi^0$ events which do not have any hadronic activity (misID~$\pi^0$), as well as the charged-current quasi-elastic (CCQE) $\nu_e$ events. For our analysis we use the simulated background rates from Ref.~\cite{deGouvea:2019wav}. We have shown the expected backgrounds per year in Fig.s \ref{fig:magneticEvents}, \ref{fig:millichargeEvents} and \ref{fig:radiusEvents} at each bin of $E_e\theta^2$ for each detector location, for CCQE (in dashed green) and misID $\pi^0$ (in dashed orange).

The DUNE flux in the (anti-)neutrino mode consists of approximately $95\%$ of $\nu_\mu~(\bar\nu_\mu)$ and $5\%$ of the wrong components, approximately $4\%$ of $\bar\nu_\mu~(\nu_\mu)$ and $1\%$ of $\nu_e+\bar\nu_e$. Therefore, we consider three different systematic uncertainties for the expected $\nu-e$ events: $10\%$ overall uncertainty in the whole beam, $1\%$ systematic uncertainty for anti-neutrinos (neutrinos) in the neutrino (anti-neutrino) beam, and $1\%$ uncertainty for the electron (anti-)neutrinos in each beam, with $f$, $f_\nu$ and $f_e$ as the corresponding pull parameters, respectively. In order to estimate how well DUNE can measure each electromagnetic property of neutrinos (described in the next section) we use the following $\chi^2$ function in each mode:
\bea\label{chi}
\chi^2&=&\sum_i\frac{\Bigg(N_i^{\rm{SM}}-N_i^{\rm{exp}}-\alpha_{CCQE}\mathbf{B}_i^{\rm CCQE}-\alpha_{\pi^0}\mathbf{B}_i^{\rm \pi^0}\Bigg)^2}{N_i^{\rm{SM}}}\nonumber\\
&+&\frac{f^2}{\sigma_f^2}+\frac{f_\nu^2}{\sigma_{f_\nu}^2}+\frac{f_e^2}{\sigma_{f_e}^2}+ \frac{\alpha^2_{CCQE}+\alpha^2_{\pi^0}}{\sigma^2_{\alpha}}\,,
\eea
where $N_i^{\rm{SM}}$ and $\mathbf{B}_i$ are the simulated SM number of events (the mock data) and the expected background at each bin of $E_e\theta^2$, respectively. We include a pull parameter ($\alpha$) for each background term with an uncertainty of $10\%$. All the $\sigma$'s are the relevant uncertainties for each pull parameter. For the neutrino mode, the expected number of events is defined as 
\bea
N_i^{\rm{exp}}(X)=(1+f)\Bigg[N_i^{\nu_\mu}(X)+f_\nu N_i^{\bar\nu_\mu}(X)+f_e N_i^{\nu_e}(X)+f_ef_\nu N_i^{\bar\nu_e}(X)\Bigg]\,,
\eea
where $X$ is the electromagnetic parameter we constrain each time. The same expression holds for the anti-neutrino beam, after changing $\nu  \leftrightarrow\bar\nu$.

\section{Electromagnetic properties of neutrinos}\label{sec:EMproperties}
\subsection{Neutrino Magnetic Moment}

\begin{figure}[t]
\includegraphics[angle=0,width=1\textwidth]{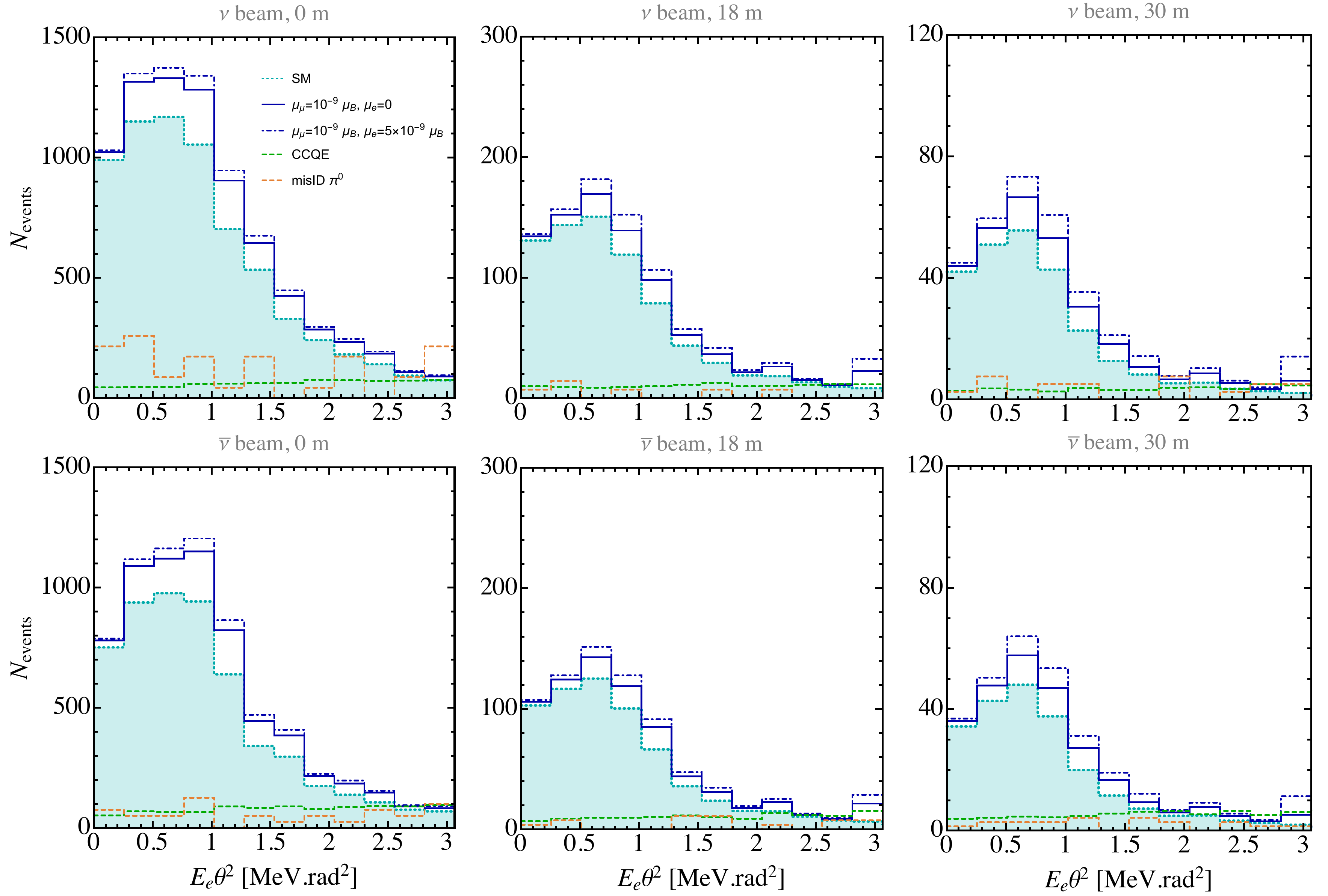}
\caption{Here we are plotting the event distributions, binned in the variable $E_{e}\theta^{2}$, for $\nu$-mode (top row) and $\bar{\nu}$-mode (bottom row) for on-axis (left column), 18 m off-axis (middle column), and 30 m off-axis (right column). 
The shaded cyan region represents the SM prediction, while the dashed light blue and dashed orange represent the CCQE and mis-identified $\pi^{0}$ backgrounds. The dark blue solid (dark blue dot-dashed) include the contributions the magnetic moment with $\mu_{\mu} = 10^{-9}~\mu_{B}$ ($\mu_{\mu} = 10^{-9}~\mu_{B}$ and $\mu_{e} = 5 \times 10^{-9}~\mu_{B}$). }
\label{fig:magneticEvents}
\end{figure}


\begin{figure}[t]
\begin{center}
\includegraphics[angle=0,width=.6\textwidth]{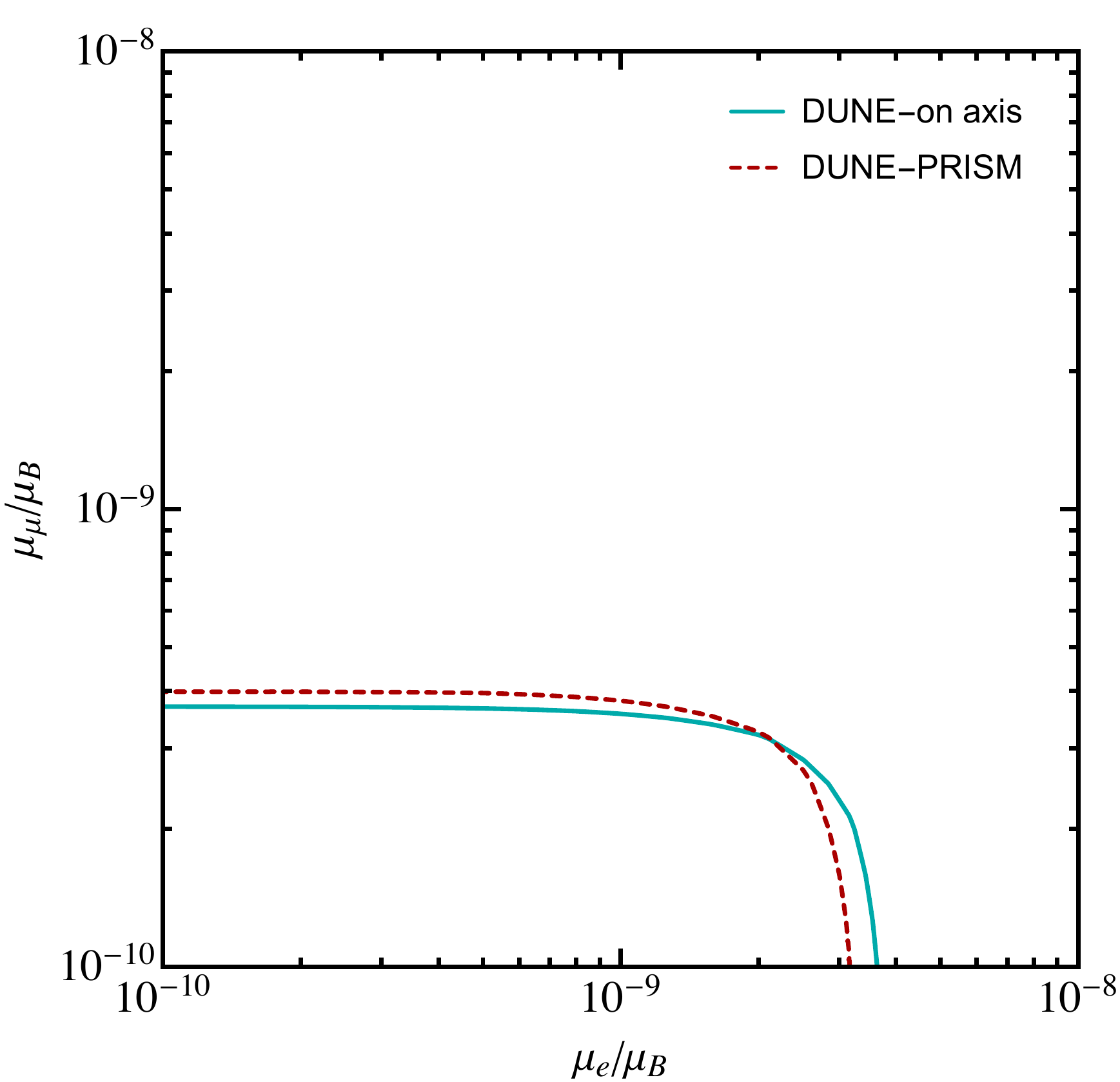}
\caption{Here we summarize our projected 90$\%$ C.L. constraints from DUNE on the neutrino magnetic dipole moment in both a fully on-axis scenario (solid cyan) as well as an off-axis DUNE-PRISM setup (dashed red). See text for details about the DUNE-PRISM assumptions. }
\label{fig:magneticAllowed}
\end{center}
\end{figure}


Although we do not have right handed neutrinos in the SM and the neutrinos masses are predicted to be zero, since neutrinos oscillate they need to be massive, and these nonzero masses imply that neutrinos have non-zero electromagnetic dipole moments.  For the Dirac neutrinos this interaction is described by the dimension-five interaction \footnote{For the Majorana neutrinos we have $\bar\nu\to \nu$.}
\be 
\mathcal{L} \supset \mu_{\alpha \beta} ~\left(\overline{\nu_{\alpha}} \sigma_{\mu\nu} \nu_{\beta} F^{\mu \nu}\right),
\ee
where the parameter $\mu$ is the magnetic moment, conventionally reported in units of the Bohr magneton, $\mu_{B} \equiv e/(2 m_{e})$. In the SM a non-zero magnetic moment can be generated at the loop level and we find the upper bound \cite{Fujikawa:1980yx}:
\be
\mu_{\alpha\beta}< \frac{3eG_f}{8\sqrt{2} \pi^2} m_\nu\ \sim 3\times10^{-20}\mu_B ~\Big(\frac{m_\nu}{0.1~ {\rm{eV}}}\Big)\,,
\ee
which is orders of magnitude far from the reach of the sensitivity of the current or near future neutrino experiments. On the other hand, many models beyond the SM predict that the neutrino has a magnetic moment (e.g.~\cite{Lindner:2017uvt,Babu:2020ivd}), hence the observation of a non-zero neutrino magnetic moment in the foreseeable future can imply the existence of physics beyond the SM.   

Neutrino-electron scattering via a magnetic moment is modified via adding the following term to the SM  cross section of Eq.~(\ref{eq:nueXSSM}) \cite{deGouvea:2006hfo}:
\be
\frac{d \sigma_{\nu_{\alpha}e}^{\rm{mm}}}{dE_{R}} = \alpha \mu_{\alpha}^{2} \left[\frac{1}{E_{R}}-\frac{1}{E_{\nu}}\right],
\ee
where $E_{R}$ is the electron recoil energy, $E_{\nu}$ is the incoming neutrino energy and $\alpha$ is the QED fine structure constant. In this work we only focus on the diagonal moments $\mu_\alpha \equiv\mu_{\alpha\alpha}$. We have shown the expected number of $\nu-e$ events in the presence of the magnetic moment term in Fig.~\ref{fig:magneticEvents}. 

The strongest terrestrial bounds on the magnetic moment include LSND~\cite{Auerbach:2001wg}, $\mu_{\nu_{\mu}} < 6.8 \times 10^{-10}$ $\mu_{B}$, and GEMMA~\cite{Beda:2009kx}, $\mu_{\nu_{e}} < 3.2 \times10^{-11}$ $\mu_{B}$. We note that Borexino~\cite{Borexino:2017fbd} has also obtained constraints from solar neutrinos: $\mu_{\nu_{e}} < 3.9 \times 10^{-11}$ $\mu_{B}$, $\mu_{\nu_{\mu}} < 5.8 \times 10^{-11}$ $\mu_{B}$, and $\mu_{\nu_{\tau}} <5.8 \times 10^{-11}$ $\mu_{B}$. 

From DUNE analysis with $3.5$ years at each mode, assuming the experiment would be done $100\%$ on axis, we find:
\bea
\mu_{\nu_\mu}<3.2\times10^{-10}~\mu_{B},\quad \mu_{\nu_e}<3.3\times10^{-9}~\mu_{B}\,,
\eea
with $90\%$ C.L.. To get these upper bounds we turn on one magnetic moment at a time. With the PRISM concept we find
\bea
\mu_{\nu_\mu}<3.4\times10^{-10}~\mu_{B},\quad \mu_{\nu_e}<2.8\times10^{-9}~\mu_{B}\,.
\eea

These results are summarized in Fig.~\ref{fig:magneticAllowed} where we show DUNE's expected sensitivity to electron- and muon-flavored magnetic moments. We note that from this it appears that DUNE will be able to place the leading terrestrial constraint on the $\nu_{\mu}$ magnetic moment by exceeding the current bound from LSND~\cite{Auerbach:2001wg}, but still an order of magnitude worse compared to Borexino. Although the PRISM can improve the bound on the electron neutrino magnetic moment, because of the lack of $\nu_e$ statistics this constraint can never compete with the existing bounds and is at least two orders of magnitude worse. 

\subsection{Neutrino Milli-Charge}

In the SM neutrinos are predicted to be electrically neutral, as a result of not having any right handed neutrinos in the Lagrangian (see e.g. the review~\cite{Giunti:2014ixa}). Once we introduce the right handed neutrinos $\nu_R$ in order to have a Dirac neutrino mass term, which is singlet under $SU(2)_L$, we are obliged to also denote a hypercharge to it, which spoils the quantization of the electric charge. As a result, not only the neutrinos will have an electric charge $q_\nu\equiv\epsilon$, but also the proton and neutron will be non-neutral: $Q_p=1-\epsilon$ and $Q_n=-\epsilon$. Obviously there are very strong constraints on the non-neutrality of neutrons, which make the electric charge of neutrinos very suppressed, ans so they are often called the neutrino milli-charge. 
\begin{figure}[t!]
\begin{center}
\includegraphics[angle=0,width=1\textwidth]{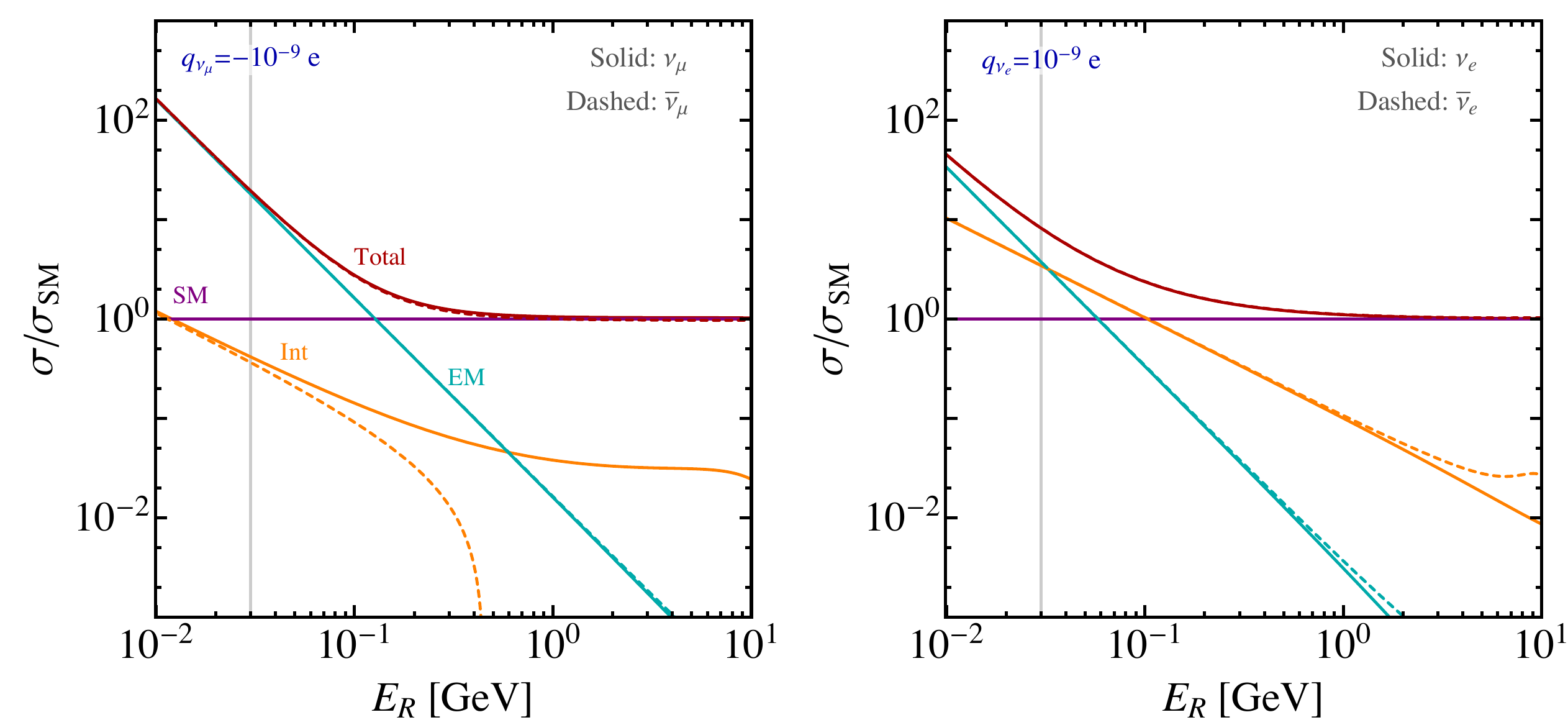}
\caption{The ratio of different terms of the milli-charge cross section over the SM cross section for the muon neutrinos (left panel) and electron neutrinos (right panel). We have shown the different contributions with the SM (purple), interference between the SM and the milli-charge contribution (Int, orange), the quadratic term with respect to the milli-charge (EM, blue) and the total cross section (red) curves. The solid (dashed) curves belong to (anti-)neutrinos. The vertical line is the 30 MeV threshold energy at DUNE and we have fixed the neutrino energy to 10 GeV}
\label{fig:mcXsec}
\end{center}
\end{figure}

\begin{figure}[t!]
\includegraphics[angle=0,width=1\textwidth]{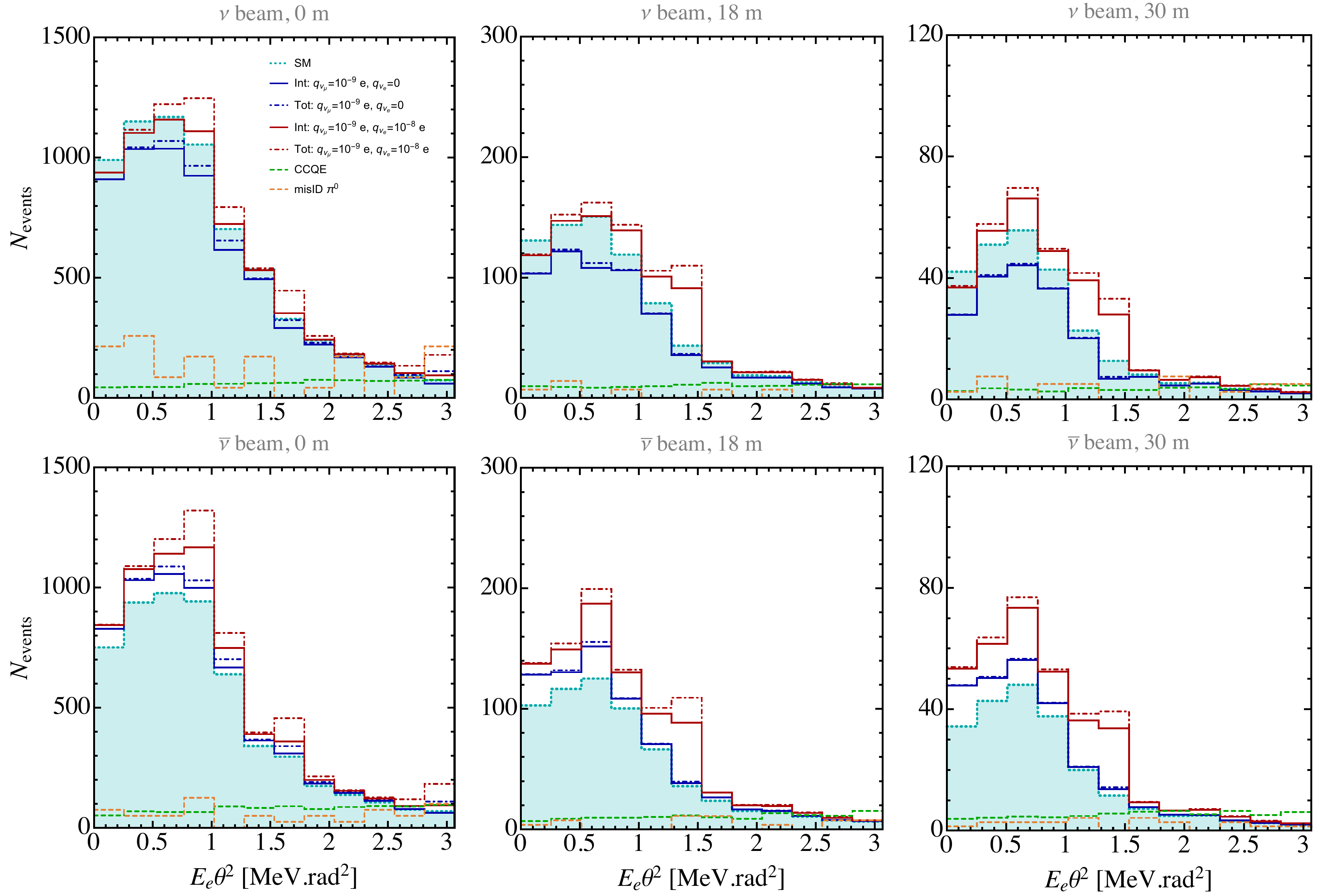}
\caption{This is the neutrino milli-charge version of Fig.~\ref{fig:magneticEvents}, with the exception that here the BSM contributions are broken into an interference (Int) and total  (Tot) contributions. The dark blue curves (dark red) correspond to milli-charge $q_{\nu_{\mu}} = 10^{-9} ~e$ ($q_{\nu_{\mu}} = 10^{-9}~ e$ and $q_{\nu_{e}} = 10^{-8} ~e$). The solid (dashed) curves denote to the interference (total) number of events. See the text for details.}
\label{fig:millichargeEvents}
\end{figure}

The milli-charge interaction is described by the following Lagrangian
\be 
\mathcal{L} \supset q_{\nu_\alpha} \overline{\nu_{\alpha}} \gamma_\mu \nu_{\alpha} A^{\mu}\,,
\ee
where $A^\mu$ is the electromagnetic field and $q_{\nu_\alpha}$ is the neutrino milli-charge. In this case the total differential $\nu-e$ cross section that takes the milli-charge contribution into account is given by
 \be\label{eq:mcXsec}
 \frac{d \sigma^{\rm{mc}}_{\nu_{\alpha}e}}{dE_{R}}=\frac{d\sigma^{\rm{SM}}_{\nu_{\alpha}e}}{dE_R}+\frac{d \sigma^{\rm{Int}}_{\nu_{\alpha}e}}{dE_{R}}+\frac{d \sigma^{\rm{EM}}_{\nu_{\alpha}e}}{dE_{R}}\,,
 \ee
 where the first term is the SM contribution given in Eq.~(\ref{eq:nueXSSM}), the second term comes from the interference between the SM and the milli-charge part of the Lagrangian and is given by:
\bea 
\frac{d \sigma^{\rm{Int}}_{\nu_{\alpha}e}}{dE_{R}} =\frac{\sqrt{2\alpha}G_f q_{\nu_\alpha}}{E_\nu^2 E_R}\left\{g_{V}^{\nu_\alpha}\Big[2E_\nu^2+E_R^2-E_R(2E_\nu+E_R)\Big]\mp g_{A}^{\nu_\alpha}\Big[E_R(2E_\nu-E_R)\Big]\right\}\,,
\eea
where the $-(+)$ sign denote to (anti-)neutrinos, the vector and axial couplings are given in Eq.~(\ref{eq:VecAxcouplings}) and we have neglected the electron mass.
In the $E_\nu\gg E_R$ limit only the first term matters and one finds $d \sigma^{\rm{Int}}_{\nu_{e}e(\bar\nu_{e}e)}/dE_{R}\propto(1+4s^2_w) q_{\nu_e}/E_R$, while $d \sigma^{\rm{Int}}_{\nu_{\mu}e (\bar\nu_{\mu}e)}/dE_{R}\propto (1-4s^2_w)q_{\nu_\mu}/E_R$ (see e.g. \cite{Parada:2019gvy}). Therefore, one can see that although for both cases the cross sections are enhanced by the smaller electron recoil energy, for the muon neutrinos there is an extra suppression due to the weak angle dependence. 
For the energies relevant to the DUNE experiment this simplification does not hold, the interference term becomes important for all neutrino flavors and becomes dominant compared to the EM term, quadratic with respect to the milli-charge, which is given below: 
\bea
\frac{d \sigma^{\rm{EM}}_{\nu_{\alpha}e}}{dE_{R}} =\pi \alpha q_{\nu_\alpha}^2\left(\frac{2E_\nu^2+E_{R}^2-2E_\nu E_{R}}{m_e E_{R}^2 E_\nu^2}\right). 
\eea
In the $E_\nu\gg E_R$ limit the above cross section simplifies to $d \sigma^{\rm{EM}}_{\nu_{\alpha}e}/dE_{R}\sim {2\pi q_{\nu_\alpha}^2}/{m_e E_R^2}$, which is the term that usually appears in the literature \cite{Studenikin:2013my,Chen:2014dsa}. We have shown the different terms of the milli-charge cross section in Fig.~\ref{fig:mcXsec}. We have also shown the expected number of $\nu-e$ events in the presence of different milli-charge contributions in Fig.~\ref{fig:millichargeEvents}.

From the neutrino experiments, reactors appear to offer the strongest constraints on an electron neutrino milli-charge with $q_{\nu_e}\lesssim 10^{-12}~e$ ~\cite{Studenikin:2013my,Chen:2014dsa}.
From the solar and the red giant cooling~\cite{Raffelt:1999gv} we find $q_{\nu}\lesssim 10^{-14}~e$, which applies to all neutrino flavors. The experimental bound on the muon neutrino is a lot weaker, with  $q_{\nu_\mu}\lesssim 10^{-8}~e$, coming from the COHERENT experiment~\cite{Cadeddu:2020lky}, which is the only existing laboratory bound on the milli-charge of the muon neutrinos.

\begin{figure}[t!]
\begin{center}
\includegraphics[angle=0,width=.6\textwidth]{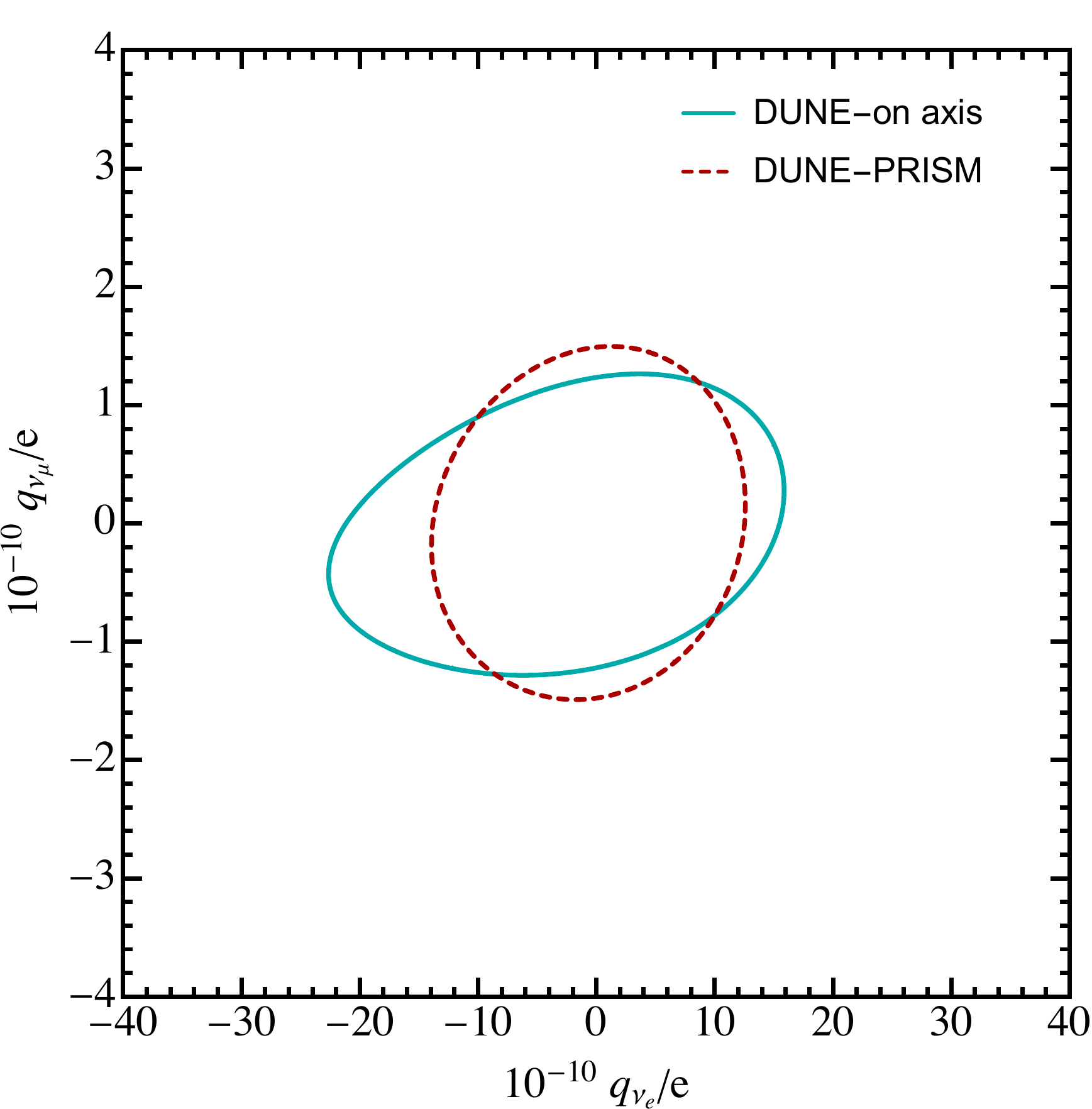}
\caption{Here we summarize our projected 90$\%$ C.L. constraints from DUNE on neutrino milli-charge in both a fully on-axis scenario (solid blue) as well as an off-axis DUNE-PRISM setup (dashed red). }
\label{fig:millichargeAllowed}
\end{center}
\end{figure}


 From DUNE analysis with $3.5$ years at each mode, assuming the experiment would be done $100\%$ on axis, we find:
\bea
-9.3\times10^{-11}~e<q_{\nu_\mu}<9.1\times10^{-11}~e,\quad -1.5\times10^{-9}~e<q_{\nu_e}<1.2\times10^{-9}~e\,
\eea
with $90\%$ C.L.. Please note that since the interference term is dominant at the range of energy of DUNE, we can also probe the sign of the milli-charge, and hence, we can report the bounds in this way. To get these upper bounds we turn on one milli-charge at a time. For the PRISM concept, we find these $90\%$ C.L. bounds:
\bea
-1.1\times10^{-10}~e<q_{\nu_\mu}<1.1\times10^{-10}~e,\quad -1.2\times10^{-9}~e<q_{\nu_e}<0.94\times10^{-9}~e\,.
\eea
Similar to the previous section, in the PRISM case we see worse bounds for the muon neutrinos, but slightly better results of the electron neutrinos. We have summerised the milli-charge results in Fig.~\ref{fig:millichargeAllowed} where we have shown the DUNE's sensitivity to $\nu_e$ and $\nu_\mu$ milli-charges. From this we see that DUNE will be able to put the leading laboratory bound on the muon neutrino milli-charge, almost two orders of magnitude better than the bound from the COHERENT experiment. 

\subsection{Neutrino Charge Radius}
Despite the magnetic moment and the milli-charge of neutrinos, their charge radii has a non-zero value even at the SM. They are induced by the radioactive corrections and are given by \cite{Bernabeu:2000hf,Bernabeu:2002pd}:
\bea
\braket{r^2_{\nu_\alpha}}_{\rm{SM}}=\frac{G_f}{4\sqrt{2}\pi^2}\Big[3-2\log \frac{m_\ell^2}{m_W^2}\Big],
\eea
where $m_\ell$ ($\ell=e,\mu,\tau$) and $m_W$ are the masses of the charged leptons and the $W^\pm$ gauge boson, respectively. Numerically this equation gives: 
\bea\label{eq:SMChargeRadii}
\braket{r^2_{\nu_e}}_{\rm{SM}}&\simeq&4.1\times10^{-33}~{\rm{cm}}^2\,,\\
\braket{r^2_{\nu_\mu}}_{\rm{SM}}&\simeq&2.4\times10^{-33}~{\rm{cm}}^2\,,\\
\braket{r^2_{\nu_\tau}}_{\rm{SM}}&\simeq&1.5\times10^{-33}~{\rm{cm}}^2\,.
\eea
Within the SM the neutrino charge radii are flavor diagonal. This is because of the conservation of the lepton numbers. Some BSM scenarios can predict off-diagonal charge radii as well (see e.g. \cite{Cadeddu:2020lky} and references there in). We ignore these off-diagonal elements in this work. 

\begin{figure}[t!]
\includegraphics[angle=0,width=1\textwidth]{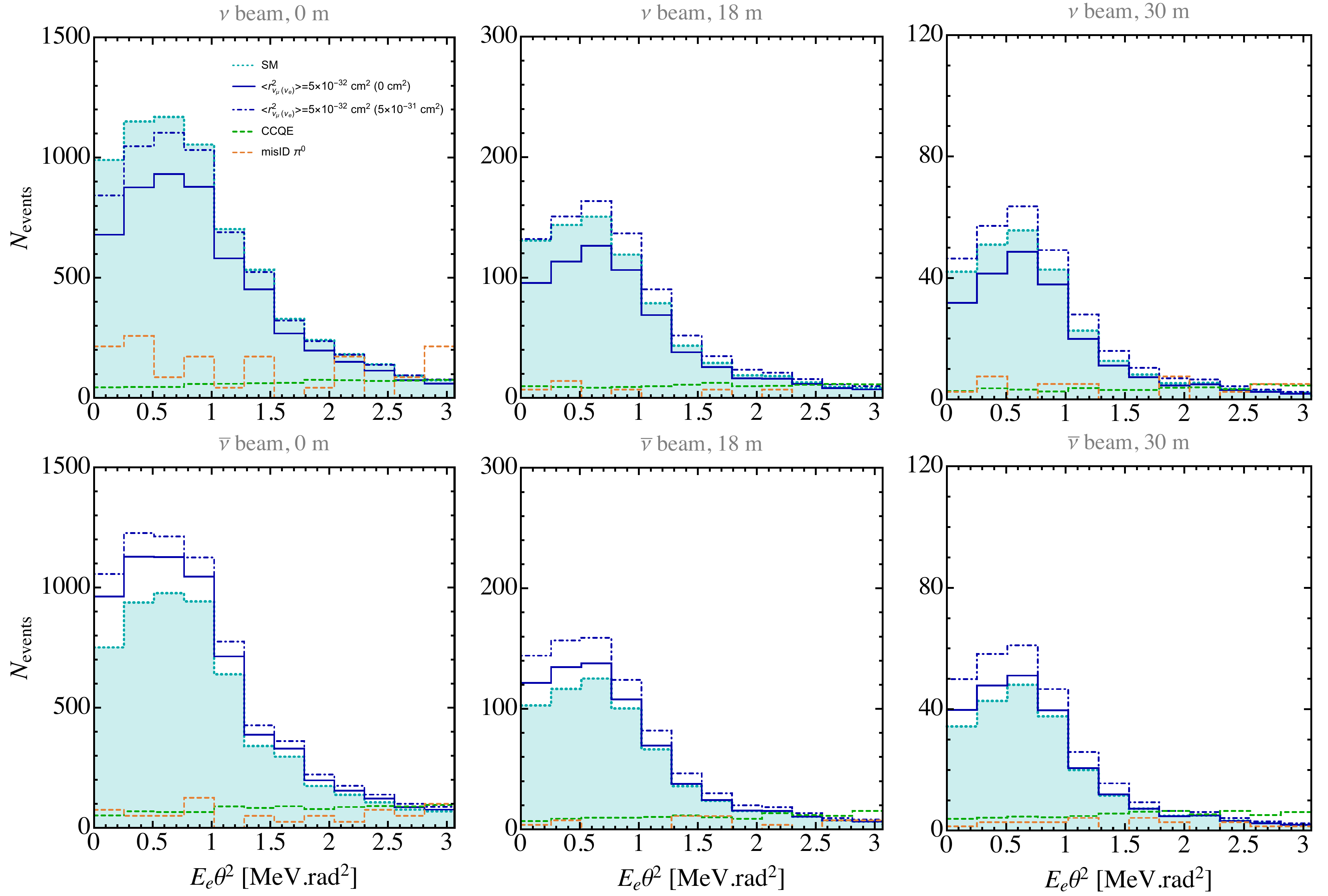}
\caption{This is the neutrino charge radius version of Fig.~\ref{fig:magneticEvents}, with the same plotting conventions.}
\label{fig:radiusEvents}
\end{figure}


\begin{figure}[t!]
\begin{center}
\includegraphics[angle=0,width=.6\textwidth]{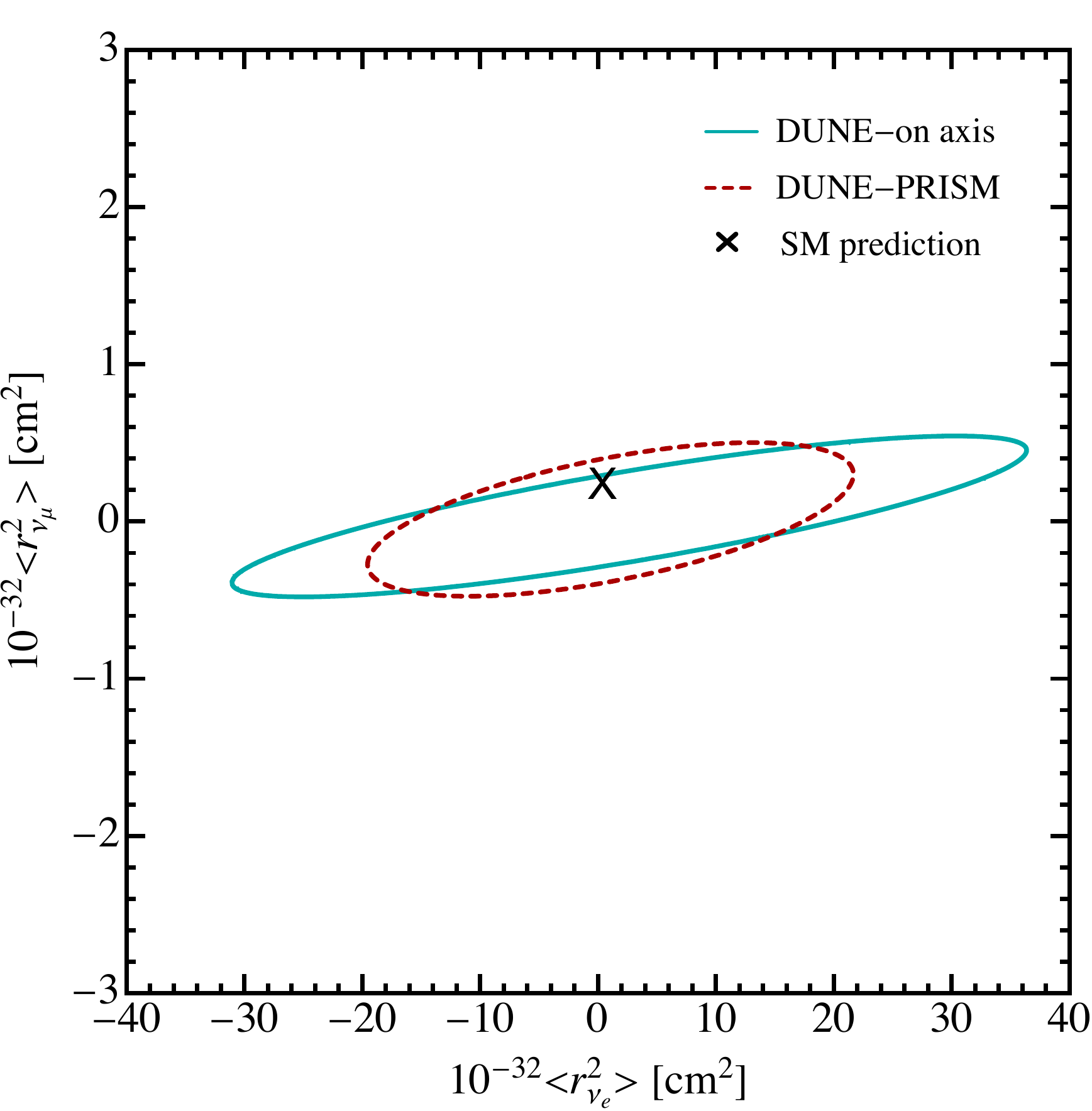}
\caption{Here we summarize our projected 90$\%$ C.L. constraints from DUNE on neutrino charge radius in both a fully on-axis scenario (solid blue) as well as an off-axis DUNE-PRISM setup (dashed red), where the black ``X'' labels the SM prediction.}
\label{fig:radiusAllowed}
\end{center}
\end{figure}


A non-zero value of the neutrinos charge radii also has an effect on their scattering on charged particles. Vogel and Engel~\cite{Vogel:1989iv} found that the modifications to neutrino scattering arising from a neutrino charge radius $\langle r^{2} \rangle $ could be written as \footnote{This differs by a factor of 2 with another convention which has been advocated in Ref.~\cite{Cadeddu:2018dux}, wherein $s^2_w \rightarrow s^2_w + \left(\frac{\sqrt{2} \pi \alpha}{3 G_{F}}\right) \langle r^{2} \rangle $. In comparing the DUNE results with the existing ones in the literature we take this difference into account.}:
\be 
g_{V}^{\nu_\alpha e} \rightarrow g_{V}^{\nu_\alpha e}+\frac{2}{3}m_W^2 \braket{r^2_{\nu_\alpha}}s^2_w\,.
\ee
Therefore, the cross section of $\nu-e$ scattering in the presence of the charge radius is the same as Eq.~(\ref{eq:nueXSSM}) after we use the above modification. Fig.~\ref{fig:radiusEvents} shows the expected number of $\nu-e$ events in the presence of the charge radius.

From a global fit of the low-energy measurements of $\nu_e-e$ and $\bar\nu_e-e$ presented in \cite{Barranco:2007ea} we find:
\bea
-2.6\times10^{-33}<\braket{r^2_{\nu_e}}<6.6\times10^{-32}\,,
\eea
assuming $90\%$ CL. The best constraint on the muon neutrino charge radius comes from a combination of the CCFR \cite{CCFR:1997zzq} and CHARM-II \cite{CHARM-II:1994aeb} data analysed in Ref.~\cite{Hirsch:2002uv}:
\bea
-5.2\times10^{-33}<\braket{r^2_{\nu_\mu}}<6.8\times10^{-33}\,.
\eea
The COHERENT bounds are also given in Ref.~\cite{Cadeddu:2020lky}, however they are an order of magnitude weaker that the current constraints. It is important to note that while the SM value of $\braket{r^2_{\nu_\mu}}$ mentioned in Eq.~(\ref{eq:SMChargeRadii}) is within the current experimental bounds, the bound on the $\nu_e$ charge radius is in tension with the SM. 

From DUNE analysis with $3.5$ years at each mode, assuming the experiment would be done $100\%$ on axis, we find:
\bea
-2.3\times10^{-33}<\braket{r^2_{\nu_\mu}}<2.2\times10^{-33},\quad -1.4\times10^{-31}<\braket{r^2_{\nu_e}}<1.4\times10^{-31}\,,
\eea
with $90\%$ C.L.. To get these upper bounds we turn on one charge radius at a time. For the PRISM case we find 
\bea
-3.2\times10^{-33}<\braket{r^2_{\nu_\mu}}<3.0\times10^{-33},\quad -1.2\times10^{-31}<\braket{r^2_{\nu_e}}<1.3\times10^{-31}\,.
\eea
We can see that using DUNE the bound on $\nu_\mu$ charge radius can be improved with a factor of two compared to the current experiments. An interesting thing to note here is that with 7-year data of DUNE at the on-axis mode it will be possible to probe the SM value of the $\nu_\mu$ charge radius for the first time, and either do a measurement or find an upper bound which can exclude the SM expectation. This result is summarised in Fig.~\ref{fig:radiusAllowed}.

\section{Conclusions}
\label{sec:concl}

We have considered the ability of the DUNE near detector to improve our understanding of neutrino properties. To examine this, we have considered three classes of electromagnetic properties: magnetic moments, milli-charges, and charge radii. 

In the case of magnetic moments, DUNE will be unable to supersede GEMMA's sensitivity to the magnetic moment of $\nu_{e}$ which is at the sub-$\mathcal{O}(10^{-11}~\mu_{B})$ level. However, our analysis has revealed that DUNE will be able to place the strongest constraints on the $\nu_{\mu}$ magnetic moment using a terrestrial neutrino source, by improving on LSND's bounds by roughly a factor of 2. We note that solar neutrino constraints from Borexino are yet stronger. 

For possible neutrino milli-charges we have been careful to include the effects of interference with the SM contributions to neutrino-electron scattering. DUNE will not be able to compete with the existing constraints on $\nu_{e}$ milli-charge from reactors. However it will be able to significantly extend experimental sensitivity to $\nu_{\mu}$ milli-charge, by improving over COHERENT's existing constraints by around two orders of magnitude. 

Lastly, the neutrino charge radius is the only EM property with a SM prediction that may be within reach of DUNE. Although the SM prediction for the $\nu_{e}$ charge radius is vastly too small to be seen at DUNE, the SM prediction for $\nu_{\mu}$ is just within its reach. 

Looking beyond the interactions considered here, a natural extension would be considering DUNE {\it inelastic} neutrino-electron scattering. One simple example would be to derive constraints on active-sterile transition magnetic moments which have recently attracted significant attention (e.g.~\cite{Gninenko:2009ks,Gninenko:2010pr,Magill:2018jla,Coloma:2017ppo,Shoemaker:2018vii,Fischer:2019fbw,Plestid:2020vqf,Jodlowski:2020vhr,Shoemaker:2020kji,Brdar:2020quo,Schwetz:2020xra,Atkinson:2021rnp,Ismail:2021dyp,Bolton:2021pey,Arguelles:2021dqn}). In this case the outgoing heavy sterile neutrino would modify the recoil spectrum compared to the diagonal magnetic moment, requiring a detailed study. 

Similarly, models including $Z'$ extensions with heavy sterile neutrinos may also induce non-trivial inelastic neutrino-electron scattering. Examples of this possibility include a dark $U(1)$ coupled to heavy SM singlet fermions~\cite{Bertuzzo:2018itn,Arguelles:2018mtc,Ballett:2019pyw} as well as gauged $U(1)_{B-L}$ extensions~\cite{Batell:2016zod}.

\vspace{1cm}

{\bf \emph{Acknowledgements-  }} ZT is thankful for the useful discussions with Pedro A.N. Machado, Joachim Kopp and Andr\'{e} de Gouv\^{e}a. The work of VM and IMS is supported by the U.S. Department of Energy under the award number DE-SC0020250. The work of ZT is supported by the Neutrino Theory Network Program Grant No. DE-AC02-07CHI11359 and the U.S. Department of Energy under the award number DE-SC0020250.

\appendix

\section{Energy and angular resolutions at DUNE}
\label{app:resolution}
In this appendix we provide details on how to calculate the expected number of $\nu-e$ events with respect to the parameter $E_e\theta^2$. We have to take into account the angular and energy resolutions of the electron measured by the experiment. We call the true parameters $E_\nu^t$, $(E_e\theta^2)^t$ and $E_e^t$, and we consider three random variables $\theta_1$, $\phi_2$ and $E_e$ which correspond to the following distribution:
\begin{eqnarray}
\rho(\phi_2,\theta_1,E_e)=\frac{1}{2\pi} N_{0,\sigma_\theta}(\theta_1)N_{E_e^t,\Sigma_{E_e}}(E_e)\,,
\end{eqnarray}
where 
\begin{equation}
    N_{a,\sigma}(s)=\frac{1}{\sqrt{2\pi}\sigma}\exp{[-\frac{(s-a)^2}{2\sigma^2}]}\,.
\end{equation}
Here $\sigma_\theta=1^{\circ}$ is the angular resolution of the outgoing electron and we have
\begin{equation}
    \Sigma_{E_e}=\sigma_{E_e}E_e\frac{1}{\sqrt{E_e/{\rm{GeV}}}}\,,
\end{equation}
where $\sigma_{E_e}=0.1$ is its energy resolution.

The next step is finding the relation between the binning parameter $(E_e\theta^2)$ that the experiment can measure and the true $(E_e^t{\theta^t}^2)$, where $\theta$ is the angle of the outgoing electron with respect to the direction of the neutrino beam, which we take $\hat{z}$. The true electron vector ${\hat{p}}^t_e=(\sin\theta_t ~\hat{x},0,\cos\theta_t~\hat{z})$ are related with
\begin{equation}
    \hat{p}_e=R_{\hat{y}}(\theta_t)R_{\hat{z}}(\phi_2)R_{\hat{y}}(\theta_1)R_{\hat{y}}(-\theta_t){\hat{p}}^t_e\,,
\end{equation}
where $R_{\hat{i}}(\alpha)$ is the rotation matrix about the axis $\hat{i}$ through the angle $\alpha$. Therefore, the measured angle $\theta$ is given by $\theta=\cos^{-1}(\hat{z}.\hat{p}_e)$.

The expected number of events at each bin of $E_e\theta^2$ with respect to the true and random variables can be found by
\begin{eqnarray}\label{eq:expEvents}
N_i=T \int d\phi_2~ d\theta_1 ~dE_e ~\rho(\phi_2,\theta_1,E_e) \times dE_\nu^t~ d(E_e\theta^2)^t ~\frac{d\phi(E_\nu^t)}{dE_\nu^t} \frac{d\sigma}{d(E_e\theta^2)^t}\,,
\end{eqnarray}
where $T=1.5\times10^{52}$ is the number of electron target particles at DUNE (assuming 50-tonnes of fiducial mass), while  $\frac{d\phi(E_\nu^t)}{dE_\nu^t}$ and $\frac{d\sigma}{d(E_e\theta^2)^t}$ are the differential flux and cross section respectively, with respect to the true parameters. We do the above integral using a Monte Carlo method. To find the number of events in each bin of the measured $E_e\theta^2$ (where $\theta$ is a function of the variables $\phi_2$ and $\theta_1$), we do as following. First we choose $n=5\times10^6$ random points for each of the parameters and we calculate the integrand of Eq.~(\ref{eq:expEvents}) for these random variables, and sum all these integrands for all the $n$ points. We call this $F_T$. Then we calculate the integrand only for the random variables which satisfy the condition $E_e\theta^2=E_e[\cos^{-1}(\hat{z}.\hat{p}_e)]^2$ in each bin $i$,  sum the result and call this $F_i$, where for the binning we have divided $0<E_e\theta^2<6m_e$ into 12 equal bins.  The number of events at each bin is then given by 
\begin{equation}
    N_i=N_T \frac{F_i}{F_T}\,,
\end{equation}
where $N_T$ is the total number of events which one can calculate without taking any of the uncertainties into account, shown in Table \ref{tab:LArrates}. It is clear that the more random variables we choose the integration will be more accurate.

\bibliography{EMDUNE}

\end{document}